\documentclass[preprintnumbers,amsmath,amssymb,prd, ,twocolumn]{revtex4}
\usepackage{graphicx}
\usepackage{amsmath,amssymb,graphics,epsfig}
\usepackage{dcolumn}
\usepackage{epsf}
\usepackage{epstopdf}
\usepackage {amssymb}
\newcommand{\nc}{\newcommand}
\nc{\ba}{\begin{eqnarray}}
\nc{\ea}{\end{eqnarray}}
\newcommand\be{\begin{equation}}
\newcommand\ee{\end{equation}}

\newcommand{\calR}{{\cal R}}

\newcommand\mPl{{M_{\rm Pl}}}

\begin{document}

\title{Violation of non-Gaussianity consistency relation
 in a single field inflationary model}

\author{Mohammad Hossein Namjoo$^{1, 3}$}
\author{Hassan Firouzjahi$^{2}$}
\author{Misao Sasaki$^{3}$}
\affiliation{$^1$School of Physics, Institute for Research in 
Fundamental Sciences (IPM),
P. O. Box 19395-5531,
Tehran, Iran}
\affiliation{$^2$School of Astronomy, Institute for Research in 
Fundamental Sciences (IPM),
P. O. Box 19395-5531,
Tehran, Iran}
\affiliation{$^3$Yukawa Institute for theoretical Physics,
 Kyoto University, Kyoto 606-8502, Japan}

\preprint{YITP-12-79, IPM/A-2012/015}

\begin{abstract}

In this paper we present a simple, toy model of single field inflation 
in which the standard non-Gaussianity consistency condition is violated. 
In this model the curvature perturbations
on super-horizon scales are not conserved and the decaying modes of 
perturbations are not negligible in the non-atractor phase.  
As a result a large local non-Gaussianity 
can be obtained in the squeezed limit which violates the standard
non-Gaussianity consistency condition for the single field models.

\end{abstract}

\maketitle
Inflation has emerged as the leading theory for the early universe and
 structure formation. There are many inflationary models which are consistent 
with recent observations.  One of the main tasks of observational cosmology is
 to constrain or rule out otherwise theoretically consistent models, reducing 
the degeneracy of the models. Meanwhile, on the theoretical side, there is an 
on-going attempt to categorize inflationary models and their predictions 
so that specific observations can exclude a class of models. One of the most 
promising classifications is single-field inflation models versus multi-field 
inflation models. Among various distinguishable observational predictions, the 
non-Gaussianity consistency relation appears to be one of the most interesting 
tests of single-field models~\cite{Maldacena:2002vr, Creminelli:2004yq},
 for a review see e.g. \cite{Koyama:2010xj,Chen:2010xka}.
It relates the amplitude of non-Gaussianity to the spectral index of the 
power spectrum in squeezed limit.
That is, in the limit $k_1 \ll k_{2}= k_{3}$ one has
\ba
\label{consistency}
\langle \calR_{\bf k_1} \calR_{\bf k_2} \calR_{\bf k_3}\rangle \simeq 
(2\pi)^3 \delta^3( \sum_i {\bf k_i} )\, (1-n_s) P_{k_1}P_{k_3}\,,
\ea
where ${\cal R}$ is the curvature perturbation on comoving surfaces, 
$n_s$ is the curvature perturbation power spectrum spectral index given by
\begin{eqnarray}
1-n_s=2\epsilon+\eta \,,
\label{specindex}
\end{eqnarray}
and $P_{k}$ is the spectrum of the comoving curvature perturbation.
The slow-roll parameters $\epsilon $ and $\eta$  are defined by
\ba
\epsilon = - \dfrac{\dot H}{H^2}\,,
\quad
\eta = \dfrac{\dot \epsilon}{H\epsilon}\, .
\label{slparas}
\ea
The relation (\ref{consistency}) has been proved explicitly by the 
effective field theory approach \cite{Cheung:2007sv} and by
a very simple, independent approach~\cite{Ganc:2010ff}. Besides that some 
physical arguments based on the fact that the curvature perturbation is 
conserved on super horizon scales for single field inflationary models
 also leads to the same result \cite{Creminelli:2004yq}. Here, however, 
we will present a counter example for this theorem. In what follows, 
we explain the model, calculate the amplitude of non-Gaussianity and 
discuss the physical reasons for the violation of the consistency relation.

Consider a canonically normalized scalar field, rolling under
 a constant potential, $V_0$. 
As usual, we assume that the energy density is dominated by the potential. 
The background evolution is given by
\ba
\label{background}
\ddot \phi +3 H\dot \phi=0\,,
\quad 3 \mPl^2 H^2 =\dfrac{1}{2} \dot \phi^2 + V_0\simeq V_0\,.
\ea
Thus we find $\dot\phi\propto a^{-3}$, and the slow roll parameters are 
given by
\ba
\label{eta}
\epsilon\propto a^{-6}\,,
\quad
\eta \simeq-6\, .
\ea
This model in the context of ultra slow-roll inflation was originally studied in \cite{Kinney:2005vj}.

The second equality shows that the absolute value of $\eta$ is always large in 
this model since $\epsilon$  is rapidly decaying. This behavior is the most 
important difference between this model and other slow-roll inflationary models.
We will see soon that this leads to interesting effects on the perturbations.
 In terms of the number of $e$-folds counted backwards from
the end of inflation,  $N \simeq H(t_e-t)\geq 0$,
one has
\ba
\label{epsilon-sol}
\epsilon(N) \simeq \epsilon_e e^{6N}\,,
\ea
where $\epsilon_e$ is the value of $\epsilon$ at the end of inflation $t=t_e$.

So far we assumed that the whole inflation is driven by 
the constant potential $V_0$. Obviously this picture has the graceful exit problem 
and we should provide a mechanism as how to terminate inflation. 
In order to overcome this problem we will slightly modify this simple 
picture at the end of this paper. We will argue that the main results do not 
change by this modification.	

As usual, the quadratic action for the 
curvature perturbation on comoving hyper-surfaces is given by
\ba
S=\dfrac{1}{2} \int d \tau d^3x\, z^2
 \left[ \calR'^2 - (\nabla \calR)^2 \right];
\quad z^2  \equiv 2\epsilon a^2 \mPl^2,
\ea   
where ${}'=d/d\tau$ and $\tau$ is the conformal time $d\tau=dt/a$.
The equation of motion of the Fourier mode is
\begin{eqnarray}
\left( a^2 \epsilon{\calR}'  \right)' + k^2a^2\epsilon \calR=0\,.
\end{eqnarray}
On super-horizon scales the solution is 
\ba
\calR= C_1+C_2 \int \frac{d\tau}{a^2\epsilon}
\ea
in which $C_1$ and $C_2$ are constants of integration.
In the conventional cases when the attractor phase has been reached, the term with $C_2$ describes a decaying mode 
which rapidly decays on super-horizon scales. However, in our case in which the system is in the non-attractor phase,  $\epsilon \propto a^{-6}$ so the would be decaying mode actually 
dominates over the constant mode.
This behavior is a signal of the violation of the consistency relation. We will see this violation by 
explicit calculations below.

Assuming the Minkowski vacuum deep 
inside the horizon, the positive frequency mode function is given by
\ba
\calR_k = C_k (-k\tau)^\nu H_\nu^{(1)}(-k\tau) \simeq 
- C_k\sqrt{\dfrac{2}{\pi}} \dfrac{1+i k \tau}{(-k\tau)^3}  e^{-i k \tau},
\ea
where 
\ba
\nu=(3+\eta)/2\simeq -\frac{3}{2} \quad , \quad
\vert C_k \vert^2 =\dfrac{\pi (-k\tau_*)^{1-2\nu}}{8 k \epsilon_* a_*^2 \mPl^2} 
\ea
and the subscript $*$ indicates the values of the parameters at 
an arbitrary reference time during inflation.
As a result the power spectrum is 
\ba
{\cal P}_\calR \equiv \dfrac{k^3}{2 \pi^2} P_k\simeq
\dfrac{H^2}{8 \pi^2 \mPl^2 \epsilon_e} 
=  \dfrac{H^2}{8 \pi^2 \mPl^2 \epsilon_k} e^{6 N_k},
\ea
where the subscript $k$ denotes the values when the mode crosses the horizon,
$a(t_k)=kH$.
Note that, due to the non-conservation of the curvature perturbation on 
super-horizon scales, the curvature perturbation must be evaluated at (or after)
the end of inflation rather than the time of horizon crossing.
The spectral index is $n_s  \simeq 4+ 2 \nu =1$, so the spectrum
is almost exactly scale invariant.

As for the bispectrum, we need the cubic action~\cite{Maldacena:2002vr},
\begin{eqnarray} \label{action3}
&&S_3=\int dt d^3x \Bigl[
a^3\epsilon^2\calR\dot{\calR}^2
+a\epsilon^2 \calR(\partial\calR)^2
\cr
&&\ -2a \epsilon \dot{\calR}(\partial
\calR)(\partial \chi)
+\frac{a^3\epsilon}{2} \dot \eta\calR^2\dot{\calR}
+\frac{\epsilon}{2a}(\partial\calR)(\partial\chi) \partial^2 \chi
\cr
&&\ +\frac{\epsilon}{4a}(\partial^2\calR)(\partial
\chi)^2
+ \left.2f(\calR)\frac{\delta L}{\delta \calR}\right|_1 \Bigr]\,,
\end{eqnarray}
where 
\begin{eqnarray} \label{solution1}
\partial^2 \chi= a^2 \epsilon\dot \calR\,,
\quad\left.\frac{\delta L}{\delta\calR}\right|_1
= a\left(\partial^2\dot\chi+H\partial^2\chi
-\epsilon\partial^2\calR \right),
\end{eqnarray}
and 
\begin{eqnarray} 
\label{redefinition}
f(\calR)&=&\frac{\eta}{4}\calR^2+\frac{1}{H}\calR\dot{\calR}
\cr
&+&
\frac{1}{4a^2H^2} \left[-(\partial\calR)(\partial\calR)
+\partial^{-2}\left(\partial_i\partial_j(\partial_i\calR\partial_j\calR) \right)
 \right] \nonumber \\
&+&
\frac{1}{2a^2H} \left[(\partial\calR)(\partial\chi)
-\partial^{-2}\left(\partial_i\partial_j(\partial_i\calR\partial_j\chi) \right)
 \right]\,.
\end{eqnarray}
The last term in cubic action can be removed by a field redefinition
$\calR \rightarrow \calR_n+f(\calR_n)$~\cite{Maldacena:2002vr}.
For this model, after field redefinition, all terms in the reduced action 
are at least of $O(\epsilon^2)$ and are negligible. 
However, $f(\calR)$ contributes to the three point function due to the field 
redefinition. The dominant contributions from $f(\calR)$ to the three point 
function comes from the first two terms in Eq.~(\ref{redefinition}). 
The remaining terms in $f(\calR)$ are suppressed on super-horizon scales by 
the spatial derivatives. Note that in the conventional analysis, such as 
in \cite{Maldacena:2002vr, Seery:2005wm, Chen:2006nt}, the second term in $f(\calR)$ is 
also discarded since $\dot {\cal R}$ is conserved on super-horizon scales.
However, in our case, we should keep this term since ${\cal R}$ is
not conserved on super-horizon scales. 

Taking into account just the first two terms  in $f(\calR)$, we end
 up with a local type non-Gaussianity (ie, squeezed limit $k_1\ll k_2=k_3$),
\ba
\langle \calR_{\bf k_1} \calR_{\bf k_2} \calR_{\bf k_3}\rangle \simeq 
(2\pi)^3 \delta^3( \sum_i {\bf k_i} )\, \frac{12}{5}f_{NL}P_{k_1}P_{k_3}\,,
\ea
with 
\ba
\label{fNL}
\frac{3}{5}f_{NL}= - \frac{3}{4} \left( \eta +4 \right)
=\dfrac{3}{2}  \, .
\ea
This apparently violates the consistency relation~(\ref{consistency}),
which would imply $(3/5)f_{NL}=\eta/4=-3/2$. As noted in the above, this is
due to the non-conservation of the comoving curvature perturbation on
super-horizon scales.

An alternative approach to the field redefinition method is to take into 
account the boundary terms that arise from the second order time 
derivatives~\cite{Arroja:2011yj}. Then we can set the terms proportional 
to $\delta L/\delta \calR|_1$ to zero, since they vanish after inserting 
the solution for the mode function. Another equivalent approach is to use 
the preliminary cubic action, before integrating by parts, in which no higher 
derivative term exists. One can check that both of these 
alternative approaches yield the same value of $f_{NL}$ as given 
in Eq.~(\ref{fNL}).
  
It is worthwhile to mention a loophole in the previous work,
which led to the consistency relation \eqref{consistency}. 
For example, by using the general action for inflation in effective field theory 
approach~\cite{Cheung:2007sv}, it is assumed that $\dot H$ is nearly a 
constant whereas in our model it decays very rapidly.

Now let us apply the $\delta N$ formalism to obtain the same results. 
Note that in general the number of $e$-folds $N$ is a function of phase space, 
$N= N(\phi, \dot \phi)$. In the conventional case, however, the slow-roll 
approximation allows us to neglect the dependence of $N$ on $\dot \phi$ and 
assume $N= N (\phi)$. On the contrary, in our case the conventional
slow-roll condition does not hold,  $\eta \simeq  -6$, so we have to consider 
$N$ as function of both $\phi$ and $\dot \phi$.
Solving the background equation of motion~\eqref{background},
we obtain
\begin{eqnarray}
\dot\phi(t)=M^2e^{-3Ht}\,,
\label{dotphisol}
\end{eqnarray}
where we assumed the sign of $\dot\phi$ to be positive without
loss of generality.
Integrating this again, we obtain
\begin{eqnarray}
\phi(t)=\frac{M^2}{3H}\left(e^{-3Ht_e}-e^{-3Ht}\right)+\phi_e\,,
\label{phisol}
\end{eqnarray}
where $\phi_e$ is the value of the scalar field at the end of
inflation, $\phi_e=\phi(t_e)$.

Combining (\ref{dotphisol}) and (\ref{phisol}),
and eliminating the integral constant $M^2$,
we obtain
\begin{eqnarray}
\phi(t)-\phi_e=\frac{\dot\phi}{3H}(e^{-3N}-1)\,,
\end{eqnarray}
Solving the above for $N$, we obtain
\begin{eqnarray}
N(\phi,\dot\phi)=\frac{1}{3}\ln
\left[\frac{\dot\phi}{\dot\phi+3H(\phi-\phi_e)}\right]\,.
\label{Nform}
\end{eqnarray}
The $\delta N$ formula is simply given by
\begin{eqnarray}
\delta N=N(\phi+\delta\phi,\dot\phi+\delta\dot\phi)-N(\phi,\dot\phi)\,.
\label{deltaN}
\end{eqnarray}

As for the fluctuations of the scalar field, one should note that unlike 
the case of the conventional models in which the constant mode dominates 
both the background field and fluctuations, in the present case 
the scalar field fluctuations are dominated 
by the constant mode (as we shall see below),
whereas both the constant and decaying modes play 
essential roles in the background, as seen from the solution~(\ref{phisol}).
In fact, this is the reason why we expressed $\delta N$ as a function of
both $\phi$ and $\dot\phi$.

Let us compute $\delta N$. According to the standard method,
we can quantize the scalar field fluctuations on the flat slices,
$\delta\phi$, which is related to the redefined field $\calR_n$
as $\delta\phi=-(\dot\phi/H)\calR_n$~\cite{Maldacena:2002vr}.
Since $V$ and $H$ are constant to a very good accuracy, we can
safely neglect the effective mass term. Hence the mode function is
exactly the same as the one for a minimally coupled massless scalar
on the fixed de Sitter background,
\begin{eqnarray}
\phi_k=\frac{H}{(2k)^{3/2}}(1+ik\tau)e^{-ik\tau}\,.
\end{eqnarray}
On super-horizon scales this can be decomposed into the
growing ($=\mbox{almost constant}$) mode and decaying mode,
\begin{eqnarray}
&&\phi_k^g\propto \cos k\tau+k\tau\sin k\tau\simeq 1\,,
\cr
\quad
&& \phi_k^d\propto k\tau\cos k\tau-\sin k\tau\simeq \frac{1}{3}(-k\tau)^3\,.
\end{eqnarray}
As a result the contributions of the decaying mode can be neglected.
Here it may be noted that the decaying mode of $\delta\phi$ corresponds
to the constant (conserved) mode of $\calR_n$,
in contrast to the conventional case.

Now we apply the above result to the $\delta N$ formula up to second order
in $\phi$ and $\dot\phi$. Since $\delta\phi$ is constant to good accuracy,
we can neglect $\delta\dot\phi$ to obtain
\begin{eqnarray}
\delta N
&\simeq&\frac{\partial N}{\partial\phi}\delta\phi
+\frac{1}{2}\frac{\partial^2 N}{\partial\phi^2}\delta\phi^2
=-\frac{H}{\dot\phi+3H(\phi-\phi_e)}\delta\phi
\cr
&&
+\frac{3H^2}{2\Bigl(\dot\phi+3H(\phi-\phi_e)\Bigr)^2}\delta\phi^2
\,.
\end{eqnarray}
We immediately find from this that $f_{NL}$ is given by
\begin{eqnarray}
\frac{3}{5}f_{NL}=+\frac{3}{2}\,.
\end{eqnarray}
This agrees with the result from the in-in formalism, Eq.~\eqref{fNL}. 
This shows that in this model the non-Gaussianity is 
generated after horizon-crossing, which we could have anticipated
because $\delta\phi$ is a free massless scalar field on the
de Sitter background in the limit $\epsilon\to0$.
Interestingly, note that had we 
neglected the contributions of the decaying mode at the background level
and assumed $N=N(\phi)$, we would have obtained $f_{NL}= -5/2$!

As mentioned before the simple picture above suffers from the graceful 
exit problem.  In order to terminate inflation suppose the potential is 
modified such that  
\ba
\label{V1}
V(\phi)=
\begin{cases}
 V_0 \qquad \quad ~\mathrm{for} \, \, \phi<\phi_c
 \\
V_2(\phi) \qquad \mathrm{for} \, \, \phi>\phi_c 
\end{cases}
\ea
in which $V_2(\phi)$ supports a second phase of slow-roll inflation 
as in conventional models. For example $V_2(\phi)$ can take the form of
a simple quadratic potential.
The value of $\phi_c$ can be fixed by requiring the continuity of the 
potential, $V_2(\phi_c)=V_0$. One then expects that the same value of 
$f_{NL}$ as in Eq.~(\ref{fNL}) should be obtained here if the modes 
relevant to CMB scales leave the horizon during the first stage of inflation.
This is a direct consequence of the fact that the curvature perturbation
on super-horizon scales freezes out during the second phase of inflation 
and there is no mechanism to change the power spectrum as well 
as the non-Gaussianity during the second phase of inflation.

For explicit calculations, note that the $\eta$ parameter is small at 
the end of the second phase of inflation, so the previously relevant 
terms are negligible here.
 However, during the transition from the first to the second 
inflationary phase, the $\eta$ parameter suddenly changes from a large
 value to nearly zero. This transition can be modeled by 
a step function $\eta = \eta_0 \left(1-\theta(t-t_c) \right)$~\cite{Arroja:2011yu},
where $\eta_0\simeq -6$ is the $\eta$ parameter at the first stage of inflation, 
and $t_c$ is the transition time at which $\phi=\phi_c$. 
This step function gives a delta function in $\dot\eta$.
Hence, the dominant term in the cubic action \eqref{action3} becomes
\ba
 S_3 \simeq \int dt \frac{a^3\epsilon}{2} \dot \eta\calR^2\dot{\calR}
 \simeq -\left[ \dfrac{a^2 \epsilon}{2} \eta_0 \calR \calR' \right]_{c}\,.
\ea
One obtains the same result as in \eqref{fNL}, 
using this term, as one should.

In conclusion, we presented a simple single-field model of inflation in which
the non-Gaussianity consistency relation is violated.
Apparently this was caused by the non-conservation of the curvature
perturbation on super-horizon scales. Our result strongly indicates that
the violation of the consistency relation occurs generically for
models in which the would-be decaying mode of the curvature perturbation
is actually dominating on super-horizon scales in the non-attractor phase. It is interesting to see
this is indeed the case and to see if there is a more realistic model
in which the violation may occur. Research in this direction is
under progress \cite{Chen:2013aj}.

\section*{Acknowledgment} 
We would like to thank F. Arroja, S. Dodelson, J-O. Gong,
L. Senatore, D. Wands and J. White for ueseful discussions and comments.
This work is supported in part by the JSPS Grant-in-Aid for Scientific Research 
(A) No.~21244033, by the YITP Overseas Exchange Program for Young Researchers,
and by MEXT Grant-in-Aid for the global COE program at 
Kyoto University, ``The Next Generation of Physics, Spun from Universality." M. H. N. would like to thank Iran's Ministry of Science and Technology for the financial supports during his visit to YITP.

{}


\begin{thebibliography}{}

\bibitem{Maldacena:2002vr} 
  J.~M.~Maldacena,
  JHEP {\bf 0305}, 013 (2003)
  [astro-ph/0210603].

\bibitem{Creminelli:2004yq} 
  P.~Creminelli and M.~Zaldarriaga,
  JCAP {\bf 0410}, 006 (2004)
  [astro-ph/0407059].

\bibitem{Koyama:2010xj} 
  K.~Koyama,
  Class.\ Quant.\ Grav.\  {\bf 27}, 124001 (2010)
  [arXiv:1002.0600 [hep-th]].

\bibitem{Chen:2010xka} 
  X.~Chen,
  Adv.\ Astron.\  {\bf 2010}, 638979 (2010)
  [arXiv:1002.1416 [astro-ph.CO]].



\bibitem{Cheung:2007sv} 
  C.~Cheung, A.~L.~Fitzpatrick, J.~Kaplan and L.~Senatore,
  JCAP {\bf 0802}, 021 (2008)
  [arXiv:0709.0295 [hep-th]].

\bibitem{Ganc:2010ff} 
  J.~Ganc and E.~Komatsu,
  JCAP {\bf 1012}, 009 (2010)
  [arXiv:1006.5457 [astro-ph.CO]].
  
\bibitem{Kinney:2005vj} 
  W.~H.~Kinney,
  Phys.\ Rev.\ D {\bf 72}, 023515 (2005)
  [gr-qc/0503017].


\bibitem{Seery:2005wm} 
  D.~Seery and J.~E.~Lidsey,
  JCAP {\bf 0506}, 003 (2005)
  [astro-ph/0503692].


\bibitem{Chen:2006nt} 
  X.~Chen, M.~-x.~Huang, S.~Kachru and G.~Shiu,
  JCAP {\bf 0701}, 002 (2007)
  [hep-th/0605045].

\bibitem{Arroja:2011yj} 
  F.~Arroja and T.~Tanaka,
  ``A note on the role of the boundary terms for the non-Gaussianity in general k-inflation,''
  JCAP {\bf 1105}, 005 (2011)
  [arXiv:1103.1102 [astro-ph.CO]].
  
\bibitem{Arroja:2011yu} 
  F.~Arroja, A.~E.~Romano and M.~Sasaki,
  Phys.\ Rev.\ D {\bf 84}, 123503 (2011)
  [arXiv:1106.5384 [astro-ph.CO]].

  
\bibitem{Chen:2013aj} 
  X.~Chen, H.~Firouzjahi, M.~H.~Namjoo and M.~Sasaki,
  arXiv:1301.5699 [hep-th].
  

\end{thebibliography}
\end{document}